\documentclass{svjour2}                    
\smartqed  
\usepackage{graphicx}
\begin{document}

\title{Entanglement and Quantum Superposition of a Macroscopic - Macroscopic
system}
\author{Francesco De Martini}

\institute{Francesco De Martini \at
              Dipartimento di Fisica, ``Sapienza'' Universit\`{a} di Roma, 
piazzale Aldo Moro 5, I-00185 Roma, Italy \\
							Accademia Nazionale dei Lincei, via della Lungara 10, I-00165 Roma,
Italy	\\
              Tel.: +39-06-49913518,\\
              Fax: +39-06-4454778\\
              \email{francesco.demartini@uniroma1.it}           }
\date{Received: date / Accepted: date}
\maketitle

\begin{abstract}
Two quantum Macro-states and their Macroscopic Quantum Superpositions (MQS)
localized in two far apart, space -- like separated sites can be non-locally
correlated by any entangled couple of single-particles having interacted in
the past. This novel \textquotedblleft Macro - Macro\textquotedblright\
paradigm is investigated on the basis of a recent study on an entangled
Micro-Macro system involving $N\approx 10^{5}$ particles. Crucial
experimental issues as the violation of Bell's inequalities by the Macro -
Macro system are considered.
\end{abstract}





The recent demonstration of the nonlocality of a Micro-Macro quantum system
composed by $N\simeq 10^{5}$ photons organized in a nearly decoherence-free
Macroscopic Quantum Superposition (MQS) \cite{DeMa08,DeMa08b,DeMa09,DeMa09b}
as well as the work done in the recent past involving photons and atoms \cite%
{Mon96,Haro96,Polz06} appear to having brought to an end the search for a
plausible realization of the famous Schr\"{o}dinger's 1935 ``paradox'' \cite%
{Sch35}. Indeed these results have finally removed any paradoxical aspect of
the ``paradox'' at the cost of raising more advanced issues e.g. regarding
the MQS\ de-coherence. As it is well known,\ a wealth of cogent and still
unresolved conceptual issues\ deeply rooted in the foundations of Quantum
Mechanics are bound to the old paradigm. First, the question regarding the
``localization'' and the ``classicality'' of the wavefunction of macroscopic
systems, i.e. the issue of ``macrorealism''\ epitomized by Einstein in a
1954 letter to Born by the sentence \textit{``Narrowness respect to the
macrocoordinates is a property not only independent of the principles of
quantum mechanics but also incompatible with them''} \cite{Eins54,DeMa09}.
Second, the intriguing interplay of the MQS dynamics with the nonlocal
correlations established with another microscopic object. This
``Micro-Macro''\ nonlocality paradigm, represented schematically in Figure
1, indeed coincides with the Schr\"{o}dinger's Cat \ (SC)\ issue, a concept
born\ in the same year 1935 in which Einstein, Podolsky and Rosen (EPR) set
forth, somewhat unwillingly the fundamental idea of nonlocality, i.e. of
the, according to Schr\"{o}dinger, ``\textit{characteristic trait of quantum
mechanics, the one that enforces its entire departure from the classical
line of thought}''\ \cite{EPR,Sch35b}.

\begin{figure}[th]
\centering
\includegraphics[width=0.55\textwidth]{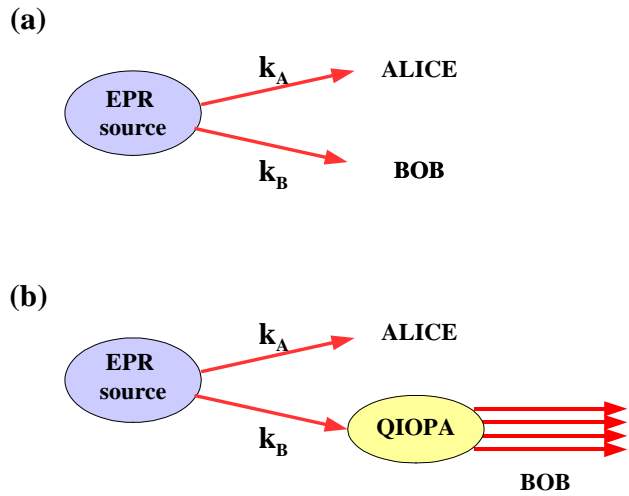}
\caption{(a)\ Generation of an entangled photon pair by spontaneous
parametric down conversion (SPDC) in a NL\ crystal: (b) Single-photon
quantum injected optical parametric amplification (QI-OPA) in a Micro -
Macro entangled configuration.}
\label{fig:fig1}
\end{figure}

The content of present work goes \textit{beyond} the SC conceptual scheme as
it ventures into the still inexplored field of ``Macro-Macro'' entanglement,
i.e. established betweem two far apart, space-like separated MQS's. By
introducing this new paradigm, and by striving for an experimental
realization we intend to complete the intriguing quantum - classical
scenario which involves some most fundamental quantum resouces, viz: EPR
nonlocality, entanglement, interference and de-coherence of Macro-states.
Aimed at this purpose and in view to our actual experimental commitment, we
believe that the simplest way to construct an overall Macro-Macro apparatus
is to combine in a smart way two of the efficient high-gain quantum-injected
optical parametric amplifiers (QI-OPA)'s succesfully adopted for the recent
Micro-Macro demonstration as well as for previous studies on phase-covariant
quantum cloning and no-signaling \cite{DeMa08,Naga07,Naga07b}. The schematic
diagram of the standard QI-OPA\ device is presented in Figure 2 and its
properties may be outlined as follows.

An entangled pair of two photons with wavelenght (wl) $\lambda $ in the
singlet state $\left| \Psi ^{-}\right\rangle _{A,B}=2^{-{\frac{1}{2}}}\left(
\left| H\right\rangle _{A}\left| V\right\rangle _{B}-\left| V\right\rangle
_{A}\left| H\right\rangle _{B}\right) \ $ was produced through a Spontaneous
Parametric Down-Conversion (SPDC)\ by the BBO nonlinear (NL)\ crystal 1
(C1)\ pumped by a pulsed UV pump beam at wl $\lambda _{P}=\lambda /2$. There 
$\left| H\right\rangle $ and $\left| V\right\rangle $ stand, respectively,
for single photon kets with horizontal $(\overrightarrow{\pi }_{H})$ and
vertical $(\overrightarrow{\pi }_{V})$ polarization $(\overrightarrow{\pi }%
)\ $while the labels $A$ (Alice) $,B$ (Bob) refer to particles associated
respectively with the spatial modes $\mathbf{k}_{A}$and $\mathbf{k}_{B}$ and
represent the two space-like separated Hilbert spaces coupled by
entanglement. The excitation source was an amplified mode-locked laser beam
frequency-doubled by second-harmonic generation and providing the OPA\
excitation field at the UV wavelength $\lambda _{P}=397.5nm$. The photon
belonging to mode $\mathbf{k}_{B}$ of each EPR pair generated by SPDC was
injected into an optical parametric amplifier consisting of a BBO NL crystal
2 (C2)\ pumped by the strong UV pump beam. A time delay secured the time
superposition in the OPA of the excitation UV pulse and of the injection
photon wavepackets. The injected single photon and the UV pump beam $\mathbf{%
k}_{P}^{\prime }$ were superimposed by means of a dichroic mirror ($DM$).
The crystal 2, cut for collinear operation along $\mathbf{k}_{B}$, emitted
over modes of linear $\overrightarrow{\pi }$, i.e.$\overrightarrow{\pi }_{H}$
and $\overrightarrow{\pi }_{V}$. The\ parametric interaction Hamiltonian $%
\widehat{H}_{B}=i\chi \hbar \widehat{a}_{H}^{\dagger }\widehat{a}%
_{V}^{\dagger }+h.c.$ acts on the single spatial mode $\mathbf{k}_{B}$ where 
$\widehat{a}_{\pi }^{\dagger }$ is the creation operator associated with $%
\overrightarrow{\pi }$. The main feature of $\widehat{H}_{B}$ is its
''phase-covariance'', i.e. invariance under $U(1)$ $\phi -$transformations,
for\ qubits $\left| \phi \right\rangle $ representing ``equatorial'' $%
\overrightarrow{\pi }-$states, i.e. $\overrightarrow{\pi }_{\phi
}=2^{-1/2}\left( \overrightarrow{\pi }_{H}+e^{i\phi }\overrightarrow{\pi }%
_{V}\right) ,\overrightarrow{\pi }_{\phi \perp }=\overrightarrow{\pi }_{\phi
}^{\perp }$, in a Poincar\'{e} sphere representation with poles: $%
\overrightarrow{\pi }_{H}$ and $\overrightarrow{\pi }_{V}$. According to the
quantum cloning theory these qubits $\left| \phi \right\rangle ,$ expressed
in terms of a single phase $\phi \in (0,2\pi )$ in the basis $\left\{ \left|
H\right\rangle ,\left| V\right\rangle \right\} $, span a privileged Hilbert
subspace corresponding to an ``optimum fidelity'', i.e. to a minimum
amplifier noise and a minimum decoherence of the amplified field \cite%
{DeMa09}. We can then re-write: $\widehat{H}_{B}=\frac{1}{2}i\chi \hbar
e^{-i\phi }\left( \widehat{a}_{\phi }^{\dagger 2}-e^{i2\phi }\widehat{a}%
_{\phi \perp }^{\dagger 2}\right) +h.c.$ where $\widehat{a}_{\phi }^{\dagger
}=2^{-1/2}(\widehat{a}_{H}^{\dagger }+e^{i\phi }\widehat{a}_{V}^{\dagger })$
and $\widehat{a}_{\phi \perp }^{\dagger }=2^{-1/2}(-e^{-i\phi }\widehat{a}%
_{H}^{\dagger }+\widehat{a}_{V}^{\dagger }) $. We shall consider the fields $%
\widehat{a}_{+}^{\dagger }$, $\widehat{a}_{-}^{\dagger }$ corresponding to $%
\phi =0$. The generic $\overrightarrow{\pi }-$state of the injected qubit $%
\left| \psi \right\rangle _{B}$ evolves into the output state $\left| \Phi
^{\psi }\right\rangle _{B}=\widehat{U_{B}}\left| \psi \right\rangle _{B}$\
according to the OPA unitary: $\widehat{U_{B}}$ =$-i\widehat{H}_{B}t/\hbar $%
. The QI-OPA\ apparatus generates in any equatorial $\overrightarrow{\pi }-$%
basis $\left\{ \overrightarrow{\pi }_{\phi },\overrightarrow{\pi }_{\phi
\perp }\right\} $, the Micro-Macro entangled state commonly referred to as
the ``SC State''\ \cite{Schl01}.

\begin{figure}[t]
\centering
\includegraphics[width=0.70\textwidth]{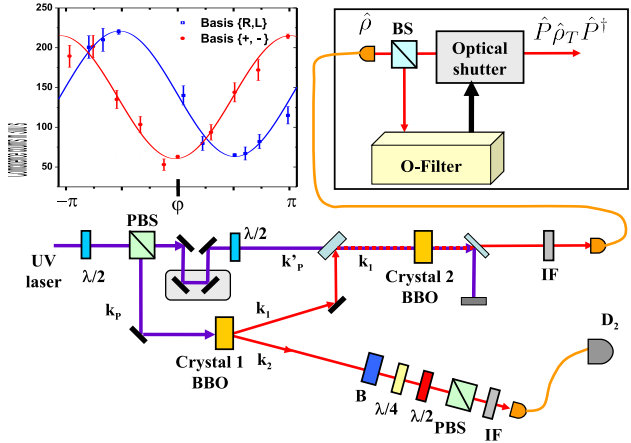}
\caption{QI-OPA setup for the Micro - Macro test.}
\label{fig:fig2}
\end{figure}

\begin{equation}
\left\vert \Sigma \right\rangle _{A,B}=2^{-1/2}\left( \left\vert \Phi ^{\phi
}\right\rangle _{B}\otimes \left\vert 1\phi ^{\perp }\right\rangle
_{A}-\left\vert \Phi ^{\phi \perp }\right\rangle _{B}\otimes \left\vert
1\phi \right\rangle _{A}\right)
\end{equation}%
where the \textquotedblright Macro-states\textquotedblright\ are: {\small 
\begin{eqnarray*}
\left\vert \Phi ^{\phi }\right\rangle _{B} &=&\sum\limits_{i,j=0}^{\infty
}\gamma _{ij}\left\vert (2i+1)\phi ;(2j)\phi ^{\perp }\right\rangle _{B} \\
\left\vert \Phi ^{\phi \perp }\right\rangle _{B}
&=&\sum\limits_{i,j=0}^{\infty }\gamma _{ij}\left\vert (2j)\phi ;(2i+1)\phi
^{\perp }\right\rangle _{B}
\end{eqnarray*}%
}with: $\gamma _{ij}\equiv C^{-2}(-\frac{\Gamma }{2})^{i}\frac{\Gamma }{2}%
^{j}\frac{\sqrt{(1+2i)!(2j)!}}{i!j!}$, $C\equiv \cosh g$, $S\equiv \sinh g$, 
$\Gamma \equiv S/C$, being $g$\ the NL\ gain. There $\left\vert p\phi ;q\phi
^{\perp }\right\rangle _{B}$ stands for a Fock state with $p$ photons with $%
\overrightarrow{\pi }_{\phi }$ and $q$ photons with $\overrightarrow{\pi }%
_{\phi \perp }$ over the mode $\mathbf{k}_{B}$. The macrostates of Eq.(2-3) $%
\left\vert \Phi ^{\phi }\right\rangle _{B}$, $\left\vert \Phi ^{\phi \perp
}\right\rangle _{B}$ are orthonormal and exhibit observables bearing
macroscopically distinct average values. In tipical experiments, under 
\textit{single particle} injection, a gain $g\simeq 4.5$ was attained
leading to a number of output photons $\mathcal{N}\simeq 5\times 10^{4}$. A
NL gain $g=6$ and a photon number $\mathcal{N}\simeq 10^{6}$ was also
attained with no substantial changes of the apparatus. Indeed, an unlimited
number of photons could be generated by the QI-OPA technique, the only
limitation being the fracture of the NL crystal in the focal region of the
laser pump. All the QI-OPA amplification properties can be greatly enhanced
by injection of \textit{multi-particle} qubits, e.g. by spin-1, 2-photon
qubits \cite{DeMa08b}. In summary, the QI-OPA\ operation acts as a perfect \ 
\textit{information preserving} \textit{quantum map} \ that transforms
nonlocally, i.e. acting between two space-like separated sites, any injected
single particle Micro-qubit, or a quantum superposition, into a
corresponding \textit{Macro-qubit }or a Macroscopic Quantum Superposition: $%
(\alpha \left\vert \phi \right\rangle _{B}+\beta \left\vert \phi ^{\bot
}\right\rangle _{B})\Longrightarrow (\alpha \left\vert \Phi ^{\phi
}\right\rangle _{B}+\beta \left\vert \Phi ^{\phi \perp }\right\rangle _{B})$%
. The two interference fringe patterns shown in left Inset of Figure 2
represent the results of a \ recent demonstration of Micro-Macro
entanglement adopting two different measurement bases.

\begin{figure}[th]
\centering
\includegraphics[width=0.80\textwidth]{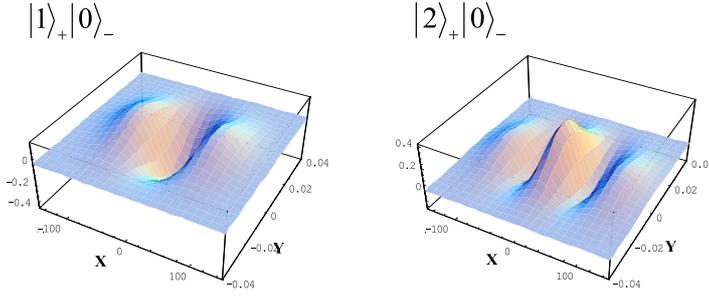}
\caption{Wigner functions of the QI-OPA amplified states with gain g = 4.
The scales on different graphs are different owing to the large squeezing
process affecting the amplified fields. The Dirac symbols at the top of the
figure represent the injection states of the OPA corresponding to the Wigner
plots below.}
\label{fig:fig3}
\end{figure}

In order to inspect at a deeper level the interference properties of our
MQSt, let's determine the Wigner function of the output field under
injection of the single-particle $\left\vert \psi \right\rangle _{B}$, by
first evaluating the \textit{symmetrically }ordered \textit{characteristic
function} of the set of complex variables $(\eta ,\eta ^{\ast },\xi ,\xi
^{\ast })\equiv \left\{ \eta ,\xi \right\} $: $\chi _{_{S}}\left\{ \eta ,\xi
\right\} \equiv \left\langle \psi \right\vert D[\eta (t)]D[\xi
(t)]\left\vert \psi \right\rangle $ expressed in terms of the \textit{%
displacement }operators $D[\eta (t)]\equiv \exp [\eta (t)\hat{a}%
_{+}(0)^{\dagger }-\eta ^{\ast }(t)\hat{a}_{+}(0)]$ and$\mathit{\ }D[\xi
(t)]\equiv \exp [\xi (t)\widehat{a_{-}}(0)^{\dagger }-\xi ^{\ast }(t)%
\widehat{a_{-}}(0)]$ where: $\eta (t)\equiv (\eta C-\eta ^{\ast }S)$; $\xi
(t)\equiv (\xi C-\xi ^{\ast }S)$. The Wigner function of the \textit{%
phase-space variables} $(\alpha ,\alpha ^{\ast },\beta ,\beta ^{\ast
})\equiv \left\{ \alpha ,\beta \right\} $ is the $4th-$dimensional Fourier
transform of $\chi _{S}\left\{ \eta ,\xi \right\} \;$\cite{Schl01}.$~$A
closed form\textit{\ }evaluation\textit{\ }of $\chi _{S}\left\{ \eta ,\xi
\right\} $ leads to the \textit{exact} expression: 
\begin{eqnarray*}
&&W(\alpha ,\beta )=\frac{1}{\pi ^{4}}\int \exp (\eta ^{\ast }\alpha -\eta
\alpha ^{\ast })\exp (\xi ^{\ast }\beta -\xi \beta ^{\ast }) \\
&&\exp [-(\left\vert \eta (t)\right\vert ^{2}+\left\vert \xi (t)\right\vert
^{2})]\left( \frac{1}{4}-\frac{\left\vert \eta (t)\right\vert ^{2}}{8}%
\right) d^{2}\eta d^{2}\xi = \\
&=&-\overline{W}(\alpha )\times \overline{W}(\beta )\times \mathcal{F(}X)
\end{eqnarray*}%
where the \textquotedblleft interference term\textquotedblright\ $\mathcal{F(%
}X)\mathcal{\ }$accounts for the quantum interference, i.e. the
superposition\ character of the MQS, of the two otherwise decoupled
quasiprobability functions: $\overline{W}(\alpha )\equiv \frac{2}{\pi }\exp
\left( -\left\vert \Delta _{A}\right\vert ^{2}\right) $ and $\overline{W}%
(\beta )\equiv \frac{2}{\pi }\exp \left( -\left\vert \Delta _{B}\right\vert
^{2}\right) $ where:$\ \Delta _{A}=\frac{1}{\sqrt{2}}\left( \gamma
_{A+}-i\gamma _{A-}\right) $, $\Delta _{B}=\frac{1}{\sqrt{2}}\left( \gamma
_{B+}-i\gamma _{B-}\right) $ and the \textit{squeezing variables} are: $%
\gamma _{A\pm }=(\alpha \pm \beta ^{\ast })e^{-g}$, $\gamma _{B\pm }=(\alpha
^{\ast }\pm \beta )e^{-g}$. The term $\mathcal{F}$ is a polynomial function
of $X\equiv \left\vert \Delta _{A}+\Delta _{B}\right\vert $\ : for 1-
particle injection is: $\mathcal{F(}X)\ $= $\left( 1-X^{2}\right) $, for
2-particle: $F(X)$= $\left( 1-2X^{2}+\frac{1}{4}X^{4}\right) $. A
3-dimensional representation of $W(\alpha ,\beta )$ for 1 and 2 photon
injection on the input mode with mode $\overrightarrow{\pi }%
_{+}=2^{-1/2}\left( \overrightarrow{\pi }_{H}+\overrightarrow{\pi }%
_{V}\right) $ and zero photon injection on $\overrightarrow{\pi }%
_{-}=2^{-1/2}\left( \overrightarrow{\pi }_{H}-\overrightarrow{\pi }%
_{V}\right) $. Note, most important, the absence of \textit{definite
positivity }of $W\left\{ \alpha ,\beta \right\} $ over the overall phase
space $\left\{ \alpha ,\beta \right\} $ which assures the quantum character
of the QI-OPA generated system \cite{Schl01,Korb05,Spekk08}.

\begin{figure}[th]
\centering
\includegraphics[width=0.60\textwidth]{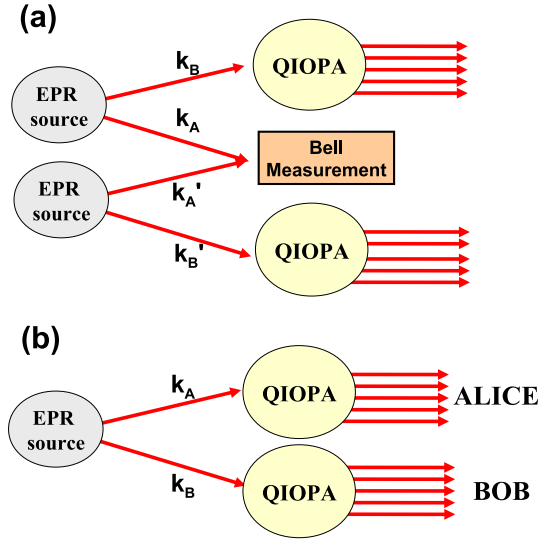}
\caption{Macro - Macro entanglement via (a) entanglement swapping (b) double
QI-OPA synchronization. }
\label{fig:fig4}
\end{figure}

In several applications the QI-OPA generated Macrostate $\rho \equiv $ $%
\left\vert \Phi \right\rangle _{B}\left\langle \Phi \right\vert $ may be
conveniently transformed, before the final measurement, by the O-Filter
device shown in Figure 2, Right Inset. This device works very generally as
follows. A beam-splitter (BS) with low reflectivity $R=(1-T)<<1$ directs
towards a subsidiary measurement apparatus a small portion of the set of
particles associated with the main beam, i.e. with the reflected state: $%
\rho _{R}\equiv tr_{T}\rho $, the trace being taken over the variables of
the \textquotedblright transmitted\textquotedblright\ field. Depending on
the measurement outcomes related to some prescribed properties of $%
\left\vert \Phi \right\rangle _{B}$, the apparatus drives, through a
suitable prescribed \textit{program }$P$, a fast electro-optical device that
may generally transforms the macrostate $\rho _{T}$= $tr_{R}\rho \approx
\rho $ emerging from the BS into the field: $(P\rho _{T}P^{+})$.\ For
instance, as shown in Figure 2 it can simply shut off, or open the path to
the main beam by a Pockels-cell fast optical-shutter, a device recently
realized and tested in our laboratory \cite{Spag08}. This subsidiary device,
denoted by the symbol OF henceforth, enables in general several interesting
quantum operations, e.g. aimed at the assessment of the \textit{%
hidden-nonlocality} of $\rho $, according to a 1995 proposal by Popescu,
widely discussed in the literature but, we believe, never realized
experimentally \cite{Pop95,Teu97,Horo09}.

The diagrams reported in Figure 4 suggest the simplest, and most convenient
arrangements of QI-OPA devices useful for our purposes. Let us discuss them
separately.

\textbf{a) QI-OPA\ entanglement-swapping}. Two independent and equal
QI-OPA's, Figure 2 (a), each possibly followed by a OF device can be
correlated, after a careful space-time sicronization of all injected and UV\
pump pulses, by a standard entanglement-swapping protocol adopting, as usual
an intermediate 50/50 beam-splitter $BS_{A}$ with input and output modes $%
(k_{A},k_{A^{\prime }})$ and $(\widetilde{k}_{A},\widetilde{k}_{A^{\prime
}}) $, respectively \cite{Pan98,Scia02}. Precisely, the overall Macrostate
at the output of the two QI-OPA\ devices connected via the Microstates
associated with the \textit{input} modes of $BS_{A}$ can be expressed as
follows: $\left\vert \Phi \right\rangle _{AB}$ = $\left\vert \Sigma
\right\rangle _{A,B}\otimes \left\vert \Sigma \right\rangle _{A^{\prime
},B^{\prime }}$ where the\ structure of two entangled Macro-states are given
by Eq.1 and by the symbol swap: $(A,B)$ $\Longrightarrow (A^{\prime
},B^{\prime })$. These states are generated by two\ equal QI-OPA's \ which
are made to interact via $BS_{A}$ through the orthogonal sets of
single-particle states $\left\{ \left\vert 1\phi ^{\perp }\right\rangle
_{A},\left\vert 1\phi ^{\perp }\right\rangle _{A^{\prime }}\right\} $ and $%
\left\{ \left\vert 1\phi \right\rangle _{A},\left\vert 1\phi \right\rangle
_{A^{\prime }}\right\} $. The overall pure state: $\left\vert \Phi
\right\rangle _{AB}\ $= $\left[ \Phi _{A}^{+}\otimes \Phi _{B}^{+}-\Phi
_{A}^{-}\otimes \Phi _{B}^{-}-\Psi _{A}^{+}\otimes \Psi _{B}^{+}+\Psi
_{A}^{-}\otimes \Psi _{B}^{-}\right] $ may be then expressed as a sum of
products of Bell states defined in the two Hilbert spaces spanned by all
eigenvectors. Precisely, the entangled Micro-states defined at the input of $%
BS_{A}$ are: $\Phi _{A}^{\pm }$=$2^{-\frac{1}{2}}\left( \left\vert 1\phi
\right\rangle _{A}\otimes \left\vert 1\phi \right\rangle _{A^{\prime }}\pm
\left\vert 1\phi ^{\perp }\right\rangle _{A}\otimes \left\vert 1\phi ^{\perp
}\right\rangle _{A^{\prime }}\right) $ and $\Psi _{A}^{\pm }$=$2^{-\frac{1}{2%
} }\left( \left\vert 1\phi \right\rangle _{A}\otimes \left\vert 1\phi
^{\perp }\right\rangle _{A^{\prime }}\pm \left\vert 1\phi ^{\perp
}\right\rangle _{A}\otimes \left\vert 1\phi \right\rangle _{A^{\prime
}}\right) $ while the Macro-states realized at the modes $\mathbf{k}_{B}$, $%
\mathbf{k}_{B^{\prime }}$ are: $\Phi _{B}^{\pm }=2^{-\frac{1}{2}}\left(
\left\vert \Phi ^{\phi }\right\rangle _{B}\otimes \left\vert \Phi ^{\phi
}\right\rangle _{B^{\prime }}\pm \left\vert \Phi ^{\phi \perp }\right\rangle
_{B}\otimes \left\vert \Phi ^{\phi \perp }\right\rangle _{B^{\prime
}}\right) $, $\Psi _{B}^{\pm }=2^{-\frac{1}{2}}\left( \left\vert \Phi ^{\phi
}\right\rangle _{B}\otimes \left\vert \Phi ^{\phi \perp }\right\rangle
_{B^{\prime }}\pm \left\vert \Phi ^{\phi \perp }\right\rangle _{B^{\prime
}}\otimes \left\vert \Phi ^{\phi }\right\rangle _{B^{\prime }}\right) $. The
expression of $\left\vert \Phi \right\rangle _{AB}$ just given shows that
the original entanglement condition existing within the two separated QI-OPA
systems $(k_{A},k_{B})$ and $(k_{A^{\prime }},k_{B^{\prime }})$ is swapped
to the ``extreme'' modes $k_{B}$ and $k_{B^{\prime }}$ by any joint Bell
single-particle measurement made on the \textit{output} modes of $BS_{A}:$ $(%
\widetilde{k}_{A},\widetilde{k}_{A^{\prime }})$. In other words, the overall
state $\left\vert \Phi \right\rangle _{AB}$ is a superposition that is
``reduced'' by any measurement taking place at the output of $BS_{A}$: e.g.,
a single-particle measurement of the Micro-Micro $\ \Phi _{A}^{+}$ leads to
a sudden reduction of $\left\vert \Phi \right\rangle _{AB}$ to the
corresponding Macro-Macro entangled state $\Phi _{E}^{+}$. This is the
relevant result\ we sought. The swapping procedure can be repeated many
times by steps involving a chain of \ BS's connected by EPR\ entangled
pairs, and lead to the concept of \ the\ \textit{quantum repeater}, a device
conceived for efficient long range communication and cryptography\ \cite%
{Brieg98,Duan01}. As far as this issue is concerned, we only remind here the
enormous \textit{degree of redundancy}, $\mathcal{N}\simeq 10^{5}$, implied
by the QI-OPA\ scheme by which the information is carried by Macro-qubits
composed by thousand or millions of particles acting simultaneously, in
parallel.

\bigskip \textbf{(b) QI-OPA\ double amplification. }This solution, shown in
Figure 4 (b), consists of the synchronous QI-OPA amplification on\ both arms
of the simple Micro-Micro EPR scheme shown in Figure 1-(a). Two independent
and equal QI-OPA amplifiers, possibly followed by two corresponding OF
devices will act independently on the EPR\ entangled modes\ $k_{A}$ and $%
k_{B}$ of a single entangled pair by the parametric evolution operators $%
\widehat{U_{A}}$ , $\widehat{U_{B}}$ corresponding to the hamiltonians $%
\widehat{H}_{A}$, $\widehat{H}_{B}$ which are formally identical but for the
swap of symbols: $A\Longrightarrow B$.\ The output Macro-state is then
simply obtained by the Micro-Macro state $\left\vert \Sigma \right\rangle
_{A,B}$, Eq.1 transformed into:\ $(\widehat{U_{A}}\otimes I)\left\vert
\Sigma \right\rangle _{A,B}=2^{-1/2}(\left\vert \Phi ^{\phi }\right\rangle
_{A}\otimes \left\vert \Phi ^{\phi \perp }\right\rangle _{B}-\left\vert \Phi
^{\phi \perp }\right\rangle _{A}\otimes \left\vert \Phi ^{\phi
}\right\rangle _{B})$, i.e. the Macro - Macro singlet we were looking for.

The assessment of the bipartite entanglement\ of $\rho $ existing betwen the
far apart sites $A$ and $B$ may be carried out by joint correlation
measurements between two standard measurement apparata located in $A$ and $B$%
, as done successfully in \cite{DeMa08} for the Micro-Macro case.\ Owing to
the\ unavoidable \textit{squeezed-vacuum noise} implied by any active
parametric amplification process, i.e. the necessary counterpart of the 
\textit{no-cloning }theorem, each measurement will be preceded by a
subsidiary state-projection transformation carried out by two OF devices
acting on the QI-OPA\ outputs states in correspondence of arms $A$, $B$. The
program $P$ \ of the OF filter may consist of the \textquotedblleft
Orthogonality Filter $\mathit{P}$\textquotedblright\ described in \cite%
{DeMa08b}. A somewhat more sophisticated procedure is requested for a
correct interpretation of \ the outcomes of any test of Bell-inequality
violation carried out at the measurement sites $A$ and $B$. According to
Popescu \textit{\textquotedblleft to prove that a quantum state is local,
one must show that the correlations between the results of any local
experiment can be described by a local hidden variable\textquotedblright } 
\textit{(LHV)\ model}\cite{Pop95}. Then, to prevent any LHV interpretation,
as suggested \ by \cite{Horo09}, the pre-measurement provided by two OF
devices should imply the adoption of corresponding OF\ programs $P$
independent of the \textit{settings} of the main mesaurement apparata
operating at $A$ and $B$. In other words, they should be once again \textit{%
phase-covariant }and, according to a recent proposal, would determine a
suitable \textquotedblleft threshold condition\textquotedblright\ to the
measured state $\rho _{T}\;$\cite{Seka09}. The structure of our experimental
apparatus includes a wide range of experimental options apt to comply with
all these requirements. Of course, our recent experience in the field tells
us that the succesful realization of the proposed experiments requires a
perfect control of all parts of the complex apparatus, a task difficult to
achieve in practice.

In conclusion, we have reported a feasible proposal that may may lead of a
relevant conceptual breakthrough in the context of some most intriguing
foundational aspects of modern Physics. Because of the conceptual
simplicity, this approach may contribute to enlighten some parts of the
elusive boundaries still existing between the ``quantum'' and the
``classical'' territories.

We acknowledge lively discussions with Pawel Horodecki, Nicolas Gisin,
Nicolo' Spagnolo, Fabio Sciarrino. The work was supported by the PRIN 2005
of MIUR and project INNESCO 2006 of CNISM.

\end{document}